# Message Dissemination in Connected Vehicles

Anirudh Paranjothi[1], Mohammed Atiquzzaman[1], and Mohammad S. Khan[2]

[1]School of Computer Science, University of Oklahoma, Norman, Oklahoma, USA
[2]Department of Computing, East Tennessee State University, Johnson City, Tennessee, USA

# Contents

# List of Figures

# List of Tables

# Part I

*Connected and Autonomous Vehicles in Smart Cities*

# 1 Message Dissemination in Connected Vehicles

Advances in connected vehicles based on Vehicular Ad-hoc Networks (VANETs) in recent years have gained significant attention in Intelligent Transport Systems (ITS) in terms of disseminating messages in an efficient manner. VANET uses Dedicated Short Range Communication (DSRC) for disseminating messages between vehicles and between infrastructures. In general, DSRC uses a dedicated 5.9 GHz band for vehicular communication [1]. DSRC has one control channel responsible for sending critical messages like information on road accidents, traffic jams, roadblocks, and six service channels responsible for sending non-critical messages like personal messages. The DSRC bandwidth is composed of eight channels that consist of six 10 MHz service channels for non-critical communications, one 10 MHz control channel for critical communications, and one 5 MHz reserved channel for future uses. As a result, VANET emerged as a promising solution for of ensuring road safety.

In a VANET environment, also known as connected vehicular environment, information is disseminated among the vehicles through messages. Two types of messages are disseminated among the vehicles: 1) periodic beacon messages and 2) event-driven messages [2, 3]. Periodic beacon messages, also known as Basic Safety Messages (BSMs) include information like vehicle's position, speed, direction, acceleration, braking status, etc. Vehicles typically broadcast BSMs to all neighboring vehicles at an interval of 100ms to 300ms. The objective of broadcasting BSMs is to be aware of the neighboring vehicles. For example, in a high dense vehicular environment like a downtown region, every vehicle receives approximately 1000 BSMs per second. Event-driven messages are generated when a vehicle encounters an event such as accidents, traffic jam, roadworks, etc. [4].

Three types of communication are possible in VANETs: 1) Vehicle to Vehicle (V2V) communication, 2) Vehicle to Infrastructure (V2I) communication , and 3) Infrastructure to Vehicle (I2V) communication . [5, 6]. V2V and V2I communication depend on DSRC for disseminating messages among the vehicles. V2V communication is used when the vehicles are in the transmission range of each other where vehicles communicate with each other using a multi-hop technique . For example, when a vehicle encounters a dangerous situation such as road accidents, loss of traction, etc., messages are disseminated to the nearby vehicles using a multi- hop technique [7, 8]. V2V communication is purely ad-hoc in nature since vehicles communicate with each other directly without any infrastructure. V2V is used for communication among vehicles. V2I is used for communicating with fixed infrastructure and long-distance communication. The components of V2I includes traffic lights, cameras, etc. V2I converts the available infrastructure into a smart infrastructure by incorporating various algorithms that use data exchange between vehicles and infrastructures. One such real-time example is Audi's "Time to green" feature that enables the vehicle to communicate with a traffic light when the traffic light is red and display the time remaining until it changes to green on the dashboard. I2V communication is most commonly used at the start of each lane of the intersection, where the fixed infrastructure such as traffic signal transmits the information of wait time to the vehicles in its transmission range . Upon receiving the traffic information, the vehicles broadcast the it to the tail end vehicles through V2V communication .

Though DSRC based communications are viable, it is still challenging to disseminate messages in a timely manner when vehicles are not in the transmission range of each other. Furthermore, DSRC communication channels are heavily congested when the vehicle density increases on the road [2, 9]. To address these limitations, two emerging paradigms: 1) vehicular cloud computing and 2) vehicular fog computing are being adopted to disseminate message between the vehicles in a connected vehicular environment. Use of cloud computing in VANETs is commonly known as vehicular cloud computing. Cloud computing has many advantages including, 1) ubiquity, 2) high processing power, and 3) location awareness , resulting in the rapid dissemination of messages among the vehicles [10]. Use of fog computing in VANETs is commonly known as vehicular fog computing, where the computations are performed at the proximity of users. Fog computing is also termed as edge computing as the computations are performed at the edge of a network [2, 11]. Vehicular fog computing uses fog nodes for the dissemination of messages among vehicles. Any real-world object can be formed as a fog node by acquiring the properties such as 1) network connectivity , 2) computation, and 3) storage .

In this book chapter, we highlight the significance of message dissemination in connected vehicles based on techniques like vehicular cloud computing, and vehicular fog computing. Our objective is to help the readers better understand the fundamentals of connected vehicles and communication techniques while disseminating messages between vehicles and between infrastructures. In a connected vehicular environment, message dissemination techniques can be modeled and performed with the help of simulations as it offers a cost-effective and scalable mechanism to analyze various parameters and scenarios. The most commonly used simulators are Simulation of Urban Mobility (SUMO) and network simulator (ns) . To simulate the trace of vehicles movements, SUMO simulator is used [12]. The output of the SUMO simulator is given as input to the ns simulator . Ns is a discrete event simulator consisting of many modules including, 1) packet loss model, 2) node deployment model [13, 14], and 3) node mobility model for dynamic network topologies to perform the simulation. For a better understanding of the simulation, we provide an overview of Hybrid-Vehcloud [15], a vehicular cloud computing based message dissemination scheme for guaranteed message delivery at high vehicle dense regions like Manhattan where buildings block radio propagation, and Dynamic Fog for Connected Vehicles (DFCV) [16], a dynamic fog computing based message dissemination scheme for rapid transmission of messages and guaranteed message delivery at high vehicle densities. In Hybrid-Vehcloud and DFCV, simulations were conducted to measure the performance of our message dissemination protocol based on the metrics like end-to-end delay of message delivery, probability of message delivery , Packet Loss

Ratio (PLR) , and average throughput . The contributions of this book chapter are as follows:

1. We provide an extensive overview of message dissemination techniques in connected vehicles including, challenges and various scenarios involved in message dissemination based on two major classifications: 1) vehicular cloud-based message dissemination, and 2) vehicular fog-based message dissemination.

2. We provide two example frameworks: Hybrid-Vehcloud, a vehicular cloud computing-based scheme, and DFCV, a fog computing-based scheme, which ensures low delay and guaranteed message delivery at high vehicle densities in a connected vehicle environment.

3. We provide a detailed analysis of the performance of various message dissemination protocols based on a defined list of performance metrics like end-to-end delay , PLR , etc.

The rest of the book chapter structured as follows: Overview of existing mes- sage dissemination techniques in VANET is illustrated in Section 1.1. The working principle and message dissemination algorithm of Hybrid-Vehcloud is presented in Section 1.2. The working principle and message dissemination algorithm of DFCV is presented in Section 1.3. Performance evaluation and simulation results of mes- sage dissemination techniques is illustrated in connected vehicles in Section 1.4. The comparison of vehicular fog computing and vehicular cloud computing is presented in Section 1.5. The conclusion and future directions of the book chapter are presented in Sections 1.6 and 1.7 respectively.

## 1.1   BACKGROUND WORK

This section provides an overview of various message dissemination techniques in connected vehicles based on two major classifications: 1) dissemination of messages using vehicular cloud computing, and 2) dissemination of messages using vehicular fog computing.

### 1.1.1   DISSEMINATION OF MESSAGES USING VEHICULAR CLOUD COMPUTING

Vehicular cloud computing used to handle complex tasks in a connected vehicular environment including, offloading large files, minimize traffic congestion, etc. One such complex task is obstacle shadowing . The radio transmissions are heavily af- fected by shadowing effects commonly known as obstacle shadowing [17]. Finding a solution for this problem plays an essential role in vehicle dense regions like a Manhattan and other downtown areas where buildings block radio wave propagation . To overcome the obstacle shadowing problem in a connected vehicle environment, an emerging technique called vehicular cloud computing is heavily being adopted by researchers and industries.

Roman et al. [18] proposed a cloud computing-based message dissemination scheme for the connected vehicular environment. The proposed scheme guarantees the integrity of messages and lower communication overhead compared to the existing message dissemination schemes. However, it has limitations such as high delay and frequent loss of connectivity . Limbasiya et al. [19] proposed a message confirmation protocol for a connected vehicular environment. The protocol helps the Road Side Units (RSU) to verify the messages obtained from other vehicles using vehicular cloud computing techniques. However, the message confirmation protocol suffers from high communication overhead and high maintenance costs . Vasudev et al. [20] illustrated a message dissemination scheme for smart transportation in vehicular cloud computing. The proposed scheme suffers from high PLR as the number of vehicles increases and thus, is not suitable for high vehicle dense regions like the downtown environment. Bi et al. [21] discussed a message dissemination pro- tocol known as Cross-Layer Broadcast Protocol (CLBP) to disseminate messages between vehicles. However, it is not suitable for vehicle-dense obstacle shadowing regions like Manhattan environment.

Syfullah et al. [22] and [10] discussed the RSU based critical message dissemination scheme to the nearby vehicles with the help of vehicular cloud networks known as Cloud-assisted Message Downlink dissemination Scheme (CMDS) and Cloud-VANET protocol respectively. CMDS and Cloud-Vanet are not suitable for high vehicle dense regions as it suffers from large transmission delays. Abbasi et al. [23] proposed a vehicular cloud-based routing algorithm for VANETs. One vehicle act as a cloud leader. The cloud leader collects information about vehicles such as position, speed, etc. through beacon messages and transmits the information to the vehicular cloud. Each vehicle has an operating system and hardware, which can provide an optimal route computed by the vehicular cloud. The cloud leader is also responsible for evaluating and monitoring the resources of other vehicles in its transmission range . However, the proposed approach suffers from high routing overhead.

Abdelatif et al. [24] presented a vehicular cloud-based traffic information dissemination approach for VANETs. The mechanism allows the vehicle to avoid highly congested and road work areas by transmitting messages like Traffic Incident Messages (TIM), event-driven messages , etc. to the vehicles present in the communication range of a vehicular cloud. But, the proposed scheme is not suitable for high dense obstacle shadowing regions. Khaliq et al. [25] proposed a vehicular data collection and data analysis scheme based on cloud computing . Once the vehicle encounters an accident, it generates an alert and transmits the information such as accident location, latitude, longitude, facing direction, etc. to the nearby vehicular cloud. Upon receiving the information, the vehicular cloud transmits the accident information to the nearest hospital and requests the ambulance service immediately. The proposed mechanism is not suitable for Manhattan regions, where the building blocks radio wave propagation from vehicles resulting in frequent disconnection of communication.

Sathyanarayanan [26] illustrated sensor-based emergency messages dissemination scheme using vehicular cloud computing. The vehicle collects information such as engine pressure, machine speed, location, etc. from the sensor at a constant time interval and reports the information to nearby gateways with a unique id and password. When the vehicle encounters an emergency scenario like a road accident, the vehicular cloud connection is established, which broadcasts the messages to all nearby vehicles in its communication range . This approach suffers from high transmission delay and routing overhead when the number of vehicles increases in the system. Mistareehi et al. [27] proposed a distributive architecture for the vehicu- lar cloud. The proposed mechanism consists of Regional Cloud (RC) and Vehicular Cloud (VC). RCs collects the information of vehicles through beacon messages and broadcast it to the vehicles in its transmission range . RCs also communicate with CC to provide a wide range of services to vehicles. For example, CC forwards the information of the stolen vehicle to RCs and vehicles in its transmission range . When a vehicle encounters a stolen car on the road, it transmits the location of the stolen car along with the license plate image to the RC. RC transmits the information of a stolen car to the police department. However, this approach suffers from high maintenance costs and communication overhead .

To overcome the limitations of message dissemination schemes [10, 18, 19, 20, 21, 22, 23, 24, 25, 26, 27], and to provide an efficient solution for obstacle shadowing problems in a connected vehicle environment, a hybrid vehicular cloud computing technique called Hybrid-Vehcloud is emerged. Hybrid-Vehcloud adopts mobile gateways (such as busses) in the vehicular cloud for messages dissemination in obstacle shadowing regions. The detailed explanation of Hybrid-Vehcloud is illustrated in Section 1.2.

### 1.1.2 DISSEMINATION OF MESSAGES USING VEHICULAR FOG COMPUTING

Fog computing , also known as edge computing is considered a new revolutionary way of thinking in wireless networking. It is an extension of cloud computing where computations are performed at the edge of the network [2]. Use of fog computing in VANET is commonly known as vehicular fog computing. Vehicular fog computing offers unique services including, location awareness , ultra-low frequency, and context information. The fog nodes can be created, deployed, and destroyed faster when compared to other traditional message dissemination techniques.

Cui et al. [28] proposed an edge computing-based scheme for message authentication in a connected vehicular environment. The part of the vehicle acts as an edge computing node and helps RSU in message authentication tasks, thus reducing the overload on RSU. However, this approach suffers from high PLR . Zhong et al. [29] illustrated a message authentication scheme known as a message authentication scheme for multiple mobile devices in intelligent connected vehicles based on edge computing for connected vehicles based on fog computing techniques . In this scheme, vehicles disseminate messages to mobile devices such as laptops, smartphones, etc. for data processing instead of sending to the cloud, thus reduce the communication overhead and delay. However, this approach has limitations such as high routing overhead and frequent loss of connectivity .

Yaqoob et al. [30] illustrated the fog assisted message dissemination scheme named Energy-Efficient Message Dissemination (E$^2$MD) for connected vehicles. In E$^2$MD, each vehicle updates status such as speed, position, and direction to the fog server. Thus, in case of critical situations, such as road accidents, the fog server informs the vehicles about road congestion and co-ordinates patrols to clear the road. The shortcomings of E$^2$MD include high maintenance cost and high delay associated with accessing and allocation of resources in the fog server.

Wang et al. [31] and Grewe et al. [32] illustrated the possibility of mobility-based fog computing for broadcasting information in a vehicular environment. However, it creates instability as the load on the fog servers increases. Noorani et al. [33] proposed a fog computing-based geographical model for VANETs. The objective of this approach is to improve data transmission and reduce the latency in inter-vehicular communications. Here, the authors considered RSUs and base stations as fog nodes responsible for transmitting data packets and providing an optimal path for routing resulting in less delay in data transmission. However, this approach suffers from high PLR .

Xiao et al. [34] presented a fog computing-based data dissemination scheme for VANETs. The roadside infrastructures with small coverage areas such as RSUs, Wi-Fi access points are considered as fog nodes , and roadside infrastructures with large coverage areas such as base stations are considered as cloud nodes. The fog nodes transmit the data received from a vehicle to all other vehicles in its transmission range . Furthermore, the vehicles can request data or available services from the fog nodes. Upon receiving the request, the fog nodes upload the request to the cloud nodes, and the requested items will be provided by the cloud nodes with a bitwise exclusive-or strategy. The shortcomings of this approach include high delays associated with accessing available services. Sarkar et al. [35] discussed the usage of fog computing techniques with the internet of things. The focus of this work is to analyze suitability and applicability fog computing in latency-sensitive applications. Also, the authors performed a comparison of the traditional cloud computing paradigm with fog computing in terms of maintenance cost and latency. From the simulation results, it is clearly depicted that the fog computing outperforms traditional cloud computing at all simulation times.

Youn [36] proposed a vehicular fog-based scheme for transmitting traffic information in a connected vehicular environment. When a vehicle encounters an accident scenario, it transmits the accident information such as location, etc. to nearby fog nodes . Fog nodes transmit the accident information to the cloud and vehicles in  its communication range . However, the proposed scheme is not suitable for high dense obstacle shadowing regions. Tang et al. [37] proposed a hierarchical fog computing model for big data analysis in smart cities. The authors analyzed the case study of a smart pipeline system and constructed a working prototype of the fog nodes to demonstrate its implementation. From the results, it is clearly depicted that fog computing architecture has significant advantages over traditional cloud architecture. However, this approach has limitations such as high routing overhead and frequent loss of connectivity .

To address the shortcomings of message dissemination schemes [28, 29, 30, 31, 32, 33, 34, 36, 37], a fog-based layered architecture, called DFCV is proposed for the dissemination of messages. DFCV uses a three-layered architecture consisting of fog computing and cloud computing techniques, thereby ensuring efficient resource utilization, rapid transmission of messages , decreases in delay and better QoS. The detailed explanation of DFCV is illustrated in Section 1.3.

## 1.2    HYBRID-VEHCLOUD MESSAGE DISSEMINATION

Hybrid-Vehcloud is a vehicular cloud computing-based scheme which ensures low delay and guaranteed message delivery at high vehicle densities in a connected vehicle environment. This section provides an overview of the Hybrid-Vehcloud message dissemination technique based on two major classifications: 1) Dissemination of messages using Hybrid-Vehcloud, 2) Hybrid-Vehcloud message dissemination algorithm.

### 1.2.1    DISSEMINATION OF MESSAGES USING HYBRID-VEHCLOUD

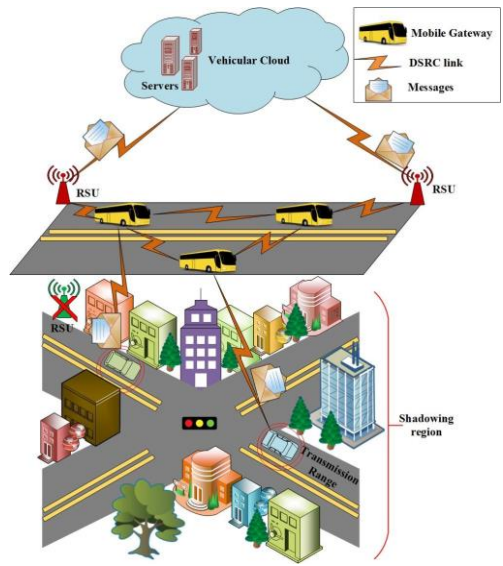

**Figure 1.1** Dissemination of messages in connected vehicles using Hybrid-Vehcloud.

The working of Hybrid-Vehcloud is illustrated with an example. Assume the vehicles represented in Fig. 1.1 are in obstacle shadowing regions and need to disseminate messages between each other. Though the vehicles are in communication range of an RSU, it is not possible to establish communication due to problems caused by shadowing . The solution for this situation is to deploy a vehicular cloud where the buses act as mobile gateways .

Following steps are involved in disseminating messages: First, vehicles collect information like the traffic jam, roadblock, etc. and disseminate them to mobile gateways using DSRC protocol. The mobile gateways deploy vehicular cloud based on infrastructure such as RSU and disseminate the message received from vehicles to the cloud. In addition, the mobile gateways transmit their own information like location, access delay, bandwidth etc. to the cloud. After receiving the information about gateways and input messages, the cloud servers assess the traffic density and determine suitable mobile gateways to disseminate the message. The gateways are selected to maximize the coverage range of vehicles in the targeted area. The messages are then transmitted to the vehicles using DSRC through an appropriate gateway. As mobile gateways are aware of the location of the vehicles probing the situation is not necessary to determine the obstacle shadowing regions . Step by step execution of Hybrid-Vehcloud algorithm is illustrated in Section 1.2.2.

## 1.2.2 HYBRID-VEHCLOUD MESSAGE DISSEMINATION ALGORITHM

---
**Algorithm 1** Hybrid-Vehcloud (input_msg)
---
1: **scan** trans_range ($V_x$)

3: **if** $V_x > 0$ **then**

4:      **for** ($i = 1$; $i <= n$; $i++$) **do**

5:          $loc[i]$ = obstacle_shadowing ($i$)

6:          **if** $loc[i] == 1$ **then**
7:              **establish** veh_cloud (input_msg)
8:              **print** message sent using Hybrid-Vehcloud technique
9:          **end if**
10:      **end for**
11: **else**

13:      **print** no nearby vehicles in obstacle shadowing regions
12: **end if**

14: **if** ($V_y == 1$) **then**
15:      **repeat** steps 1 to 13
16: **end if**

---

The Hybrid-Vehcloud algorithm works as follows: First, the set of neighboring vehicles in the transmission range of a base station associated with a sender is calculated. Then, trans_range ($V_x$) is used to discover the number of vehicles in the range of the base station. If the number of number of vehicles is greater than zero, the loca- tion of vehicles is determined using the obstacle_shadowing() function. The value 1 represents the vehicles is in a shadowed region and hence the messages are broadcasted using the vehicular cloud technique. The notations used in Hybrid-Vehcloud are illustrated in Table 1.1.

**Table 1.1**

**Notations used in Hybrid-Vehcloud algorithm**

| Variables | Purpose |
|---|---|
| $V_x$ | Vehicle broadcasts the message (sender) |
| $n$ | Number of vehicles in the transmission range of a sender (i.e., receiver(s)) |
| $loc$ | Either 0 or 1, based on this message the dissemination technique is determined |
| $V_y$ | New vehicle enters the transmission range of base station associated with a sender |

## 1.3 DFCV MESSAGE DISSEMINATION

DFCV is a dynamic fog computing based message dissemination scheme for rapid transmission of messages and guaranteed message delivery at high vehicle densities. This section provides an overview of the DFCV message dissemination technique based on two major classifications: 1) Dissemination of messages using DFCV, 2) DFCV message dissemination algorithm.

### 1.3.1 DFCV MESSAGE DISSEMINATION TECHNIQUE

Vehicular fog-based layered architecture, called DFCV, is shown in Fig. 1.2. DFCV consists of three layers: 1) Terminal layer, 2) Fog layer , and 3) Cloud layer.

**Terminal Layer:** This layer closest to the physical environment and end user. It consists of various devices like smartphones, vehicles, sensors, etc. As the motive of DFCV approach is to broadcast the messages in a connected vehicular environment, only vehicles are represented in the terminal layer. Moreover, they are responsible for sensing the surrounding environment and transmitting the data to the fog layer for processing and storage .

**Fog Layer:** Fog layer is located at the edge of a network. It consists of fog nodes , which includes access points, gateways, RSUs, base station, etc. In DFCV, RSUs and base stations play a major role in disseminating the messages. Fog layer can be static at a fixed location or mobile on moving carriers such as in the vehicular environment. Also, they are responsible for processing the information received from the terminal device and temporarily store it or broadcast over the network.

**Cloud Layer:** The main function of the cloud computing in DFCV is to keep track of the resources allocated to each fog node and to manage interaction and interconnection among workloads on a fog layer .

DFCV incorporates all possible scenarios for disseminating the messages including, fog-split and fog-merge . Fog split will occur in two scenarios: either the capacity of the DFCV is greater than the pre-defined threshold capacity , or the distance between the vehicle increases from the view of the sender, also known as the first ob-server. Fog merge will occur in two scenarios: either the capacity of the DFCV is less than the pre-defined threshold capacity , or the distance between the vehicles is less than the minimum distance. Step by step execution of DFCV algorithm is illustrated in Section 1.3.2.

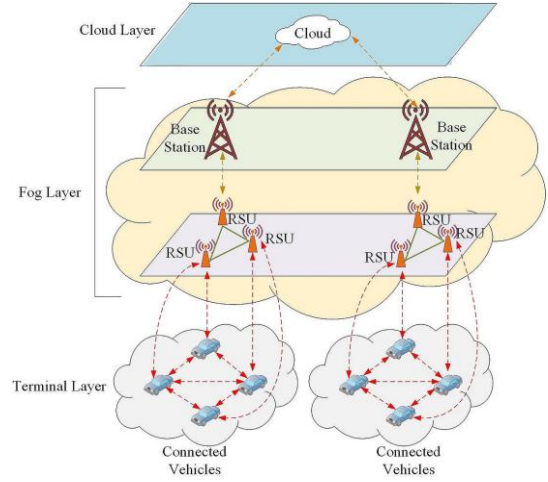

**Figure 1.2** Dissemination of messages in connected vehicles using DFCV.

### 1.3.2 DFCV MESSAGE DISSEMINATION ALGORITHM

---

**Algorithm 2** DFCV (input_msg, $veh_{send}$ , $veh_{rec}$)

---

1: **for** $veh_{send}$ $\in bs_i$ **do**
2:     **for** $v \in c$ **do**
3:         **calculate** distance()
4:         **if** ($distance > d_{min}$ f$_l$||$> th_{cap}$) **then**
5:             **split** ($v \in c$)
6:             fog_layer = $v$
7:             $bs_i$ = send (input_msg)
8:             $veh_{rec}$ = $bs_i$
9:         **else**
10:            **merge** ($v \in c$)
11:            fog_layer = $v$
12:            $bs_i$ = send (input_msg)
13:            $veh_{rec}$ = $bs_i$
14:         **end if**
15:         **print** message sent using DFCV technique
16:     **end for**
17: **end for**

---

DFCV aims to transmit the messages to the neighboring vehicles using fog computing technique . It mainly concentrates on merge and split scenarios as discussed in Section 1.3.1. The split is a primitive operation performed by DFCV using split() function. The steps are as follows: First, the distance between the vehicles is calculated using the distance() function. It is calculated based on the distance from the sender, and then, the capacity of the DFCV is determined using th_cap. The split accomplished when the distance exceeds the mini-mum distance ($d_{min}$) or the capacity of the DFCV ($f_c$) surpass the threshold capacity . Here, a single fog will split into two parts. After the split, messages are relayed to the base station with the help of the RSU and send() function is used to send the input message to the vehicles   in a corresponding base station ($bs_i$). The notations used in DFCV are illustrated in Table 1.2. Merge is another primitive operation performed by DFCV using merge() function based on the following constraints: the distance is lesser than the minimum distance ($d_{min}$), or the capacity of the DFCV ($f_c$) is lesser than the threshold capacity . It combines two or more fog layers under the same base station ($bs_i$) into a single fog layer. Then, the messages are broadcasted to the neighboring vehicle using send() function.

**Table 1.2**
**Notations used in DFCV algorithm**

| Variables | Purpose |
| --- | --- |
| $veh_{send}$ | Set of vehicle(s) that need to transmit messages |
| $veh_{rec}$ | Intended recipient(s) |
| $th$ | Threshold capacity of the fog |
| $bs_i$ | Base station associated with vehsend |

DFCV aims to provide rapid transmission of messages and guaranteed message delivery at high vehicle densities. In a connected vehicular environment, many challenges still exist due to the difficulties in the deployment and management of resources. In specific, the current techniques for V2V and V2I communications do not provide guaranteed message delivery resulting in messages being dropped be- fore reaching the destination. It is due to an instability of DSRC , arising from the frequency band used by DSRC , as the number of vehicles increases. Furthermore, the current techniques for message dissemination have limitations such as the efficient utilization of resources, delay constraints due to high mobility and unreliable connectivity, and Quality of Service (QoS). DFCV message dissemination technique addresses the shortcomings of V2V and V2I communications and broadcasts messages to the vehicles using the DFCV message dissemination algorithm. DFCV algorithm can be used in both highway and urban environments as it provides better performance compared to Hybrid-Vehcloud at all vehicle densities. However, DFCV suffers from shadowing effects caused by obstacle shadowing regions.

The objective of the Hybrid-Vehcloud message dissemination technique is to lower the delay and to provide guaranteed message delivery in obstacle shadow- ing regions at high vehicle densities in a connected vehicular environment. The radio transmissions are heavily affected by shadowing effects caused by obstacles like tall buildings, skyscrapers, etc. To overcome the shadowing effects , the Hybrid-Vehcloud message dissemination algorithm adopts a vehicular cloud for broadcasting the messages in obstacle shadowing regions , where the buses act as mobile gateways . Hybrid-Vehcloud message dissemination algorithm can be used in obstacle regions such as Manhattan and other downtown regions. Hybrid-Vehcloud performs better compared to DFCV in obstacle shadowing regions.

## 1.4   PERFORMANCE EVALUATION

In a connected vehicular environment, message dissemination techniques including, Hybrid-Vehcloud and DFCV are evaluated using simulations. The most commonly used simulators are SUMO and ns . In this section, we discuss the simulation setup and the most commonly used metrics involved in the simulation of existing message dissemination techniques.

### 1.4.1   SIMULATION SETUP

Simulations of the message dissemination protocols performed based on the algo- rithms. Algorithms illustrate step by step execution of the appropriate message dis- semination framework. For example, Hybrid-Vehcloud and DFCV perform simula- tions based on the algorithms discussed in Section 1.2.2. and 1.3.2. To simulate the trace of vehicle movements, the SUMO simulator is used [38]. The output files of the SUMO simulations are usually in the .xml format, contain information such as vehi- cle position, trip information, vehicle routes, etc. The output of the SUMO simulator is given as input to the ns simulator. Ns is a discrete event simulator, provides sub- stantial support for simulation of wired and wireless networks [39]. Ns simulator out- put trace file for every simulation. From the trace files, simulation data are collected and converted into graphs, represented in Fig. 1.3. The simulation files are publicly available for everyone in our GitHub repository [40]. For a better understanding of the simulations, we provide extensive simulation results of Hybrid-Vehcloud and DFCV in Section 1.4.3. and Section 1.4.4. The most important parameters used in simulation are represented in Table 1.3.

### 1.4.2   PERFORMANCE METRICS

The most important performance metrics in the existing message dissemination pro- tocols are:

1. End-to-end delay : Time is taken for a message to be disseminated across a network from source to destination.

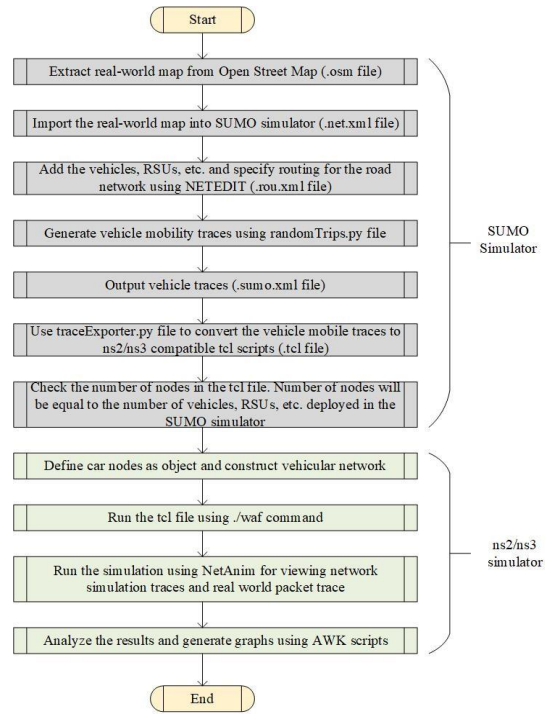

**Figure 1.3** Simulation Workflow of SUMO and ns2/ns3 simulators.

**Table 1.3**
**Most important parameters used in simulation**

| Parameters | Value |
| --- | --- |
| Road length | 10 km |
| Number of vehicles/nodes | 50-450 |
| Vehicle speed | 30-60 mph |
| Transmission range | 300 m |
| Message size | 256 bytes |
| Simulator used | ns-2, ns-3, SUMO |
| Data rate | 2 Mbps |
| Technique used | Vehicular fog, Vehicular cloud |
| Protocol | IEEE802.11p |

2. The probability of message delivery : The probability of the input message delivered to the targeted vehicles.
3. PLR : The ratio of number of lost packets to the total number of packets sent across a network.
4. Average throughput : Average rate of successfully disseminated messages across a communication channel.

### 1.4.3   PERFORMANCE EVALUATION OF HYBRID-VEHCLOUD

Hybrid-Vehcloud is compared with three previous cloud-based message dissemination schemes: 1) CMDS [22], 2) CLBP [21], and 3) Cloud-VANET protocols [10]. The results are discussed below:

**End-to-end delay:** In Hybrid-Vehcloud, knowledge of nearby vehicles significantly reduces the route setup time and propagation time across a network. Hence, it delivers the message much faster when compared to CLBP, CMDS, and Cloud-VANET protocols, represented in Fig. 1.4. The end-to-end delay is calculated against the number of vehicles, and it increases as the number of users increases in the system due to a large number of messages need to be delivered within a specific time interval.

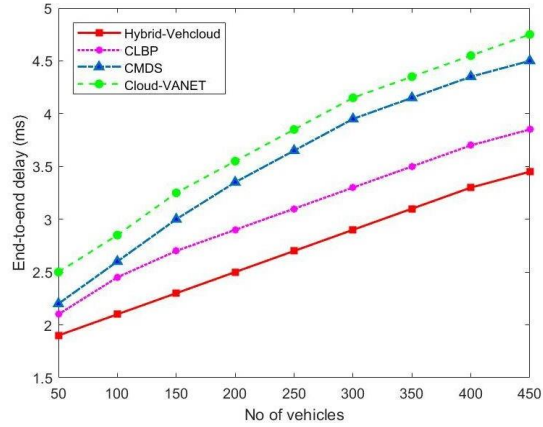

**Figure 1.4** Comparison of end-to-end delay of Hybrid-Vehcloud with CLBP, CMDS, and Cloud-VANET approaches.

**Probability of message delivery:** The probability of message delivery of Hybrid-VehCloud was observed to be higher due to guaranteed message delivery to the vehicles in the obstacle shadowing region, represented in Fig. 1.5. Probability of message delivery is calculated against the number of vehicles. For each user, the probability of message delivery distributed in the range of (0-1). From Fig. 1.5, we can observe that the probability of message delivery decreases marginally as the number of users increases due to the increase in load on Hybrid-Vehcloud.

**PLR:** The PLR of Hybrid-Vehcloud is calculated against the number of vehicles,

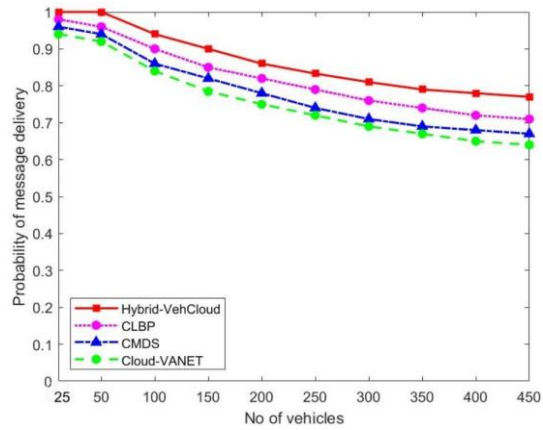

**Figure 1.5** Comparison of probability of message delivery of Hybrid-Vehcloud with CLBP, CMDS, and Cloud-VANET approaches.

represented in Fig. 1.6. PLR increases marginally as the number of vehicles increases due to the high channel congestion and frequent loss of connectivity . However, PLR of Hybrid-Vehcloud observed to be lower compared to CLBP, CMDS, and Cloud-Vanet protocols at all vehicle densities.

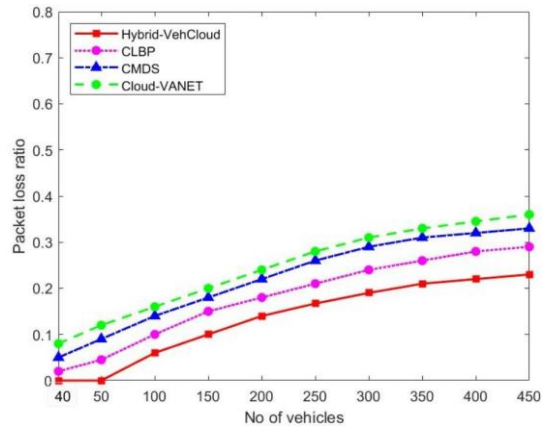

**Figure 1.6** Comparison of packet loss ratio of Hybrid-Vehcloud with CLBP, CMDS, and Cloud-VANET approaches.

**Average Throughput:** In Hybrid-Vehcloud, average throughput is the number of messages disseminated across a communication channel. From Fig. 1.7, it can be observed that average throughput increases as the number of vehicle increases in the system, due to a large number of messages disseminated between vehicles. The average throughput of Hybrid-Vehcloud is high when compared to CLBP, CMDS, and Cloud-VANET protocols at all vehicle densities.

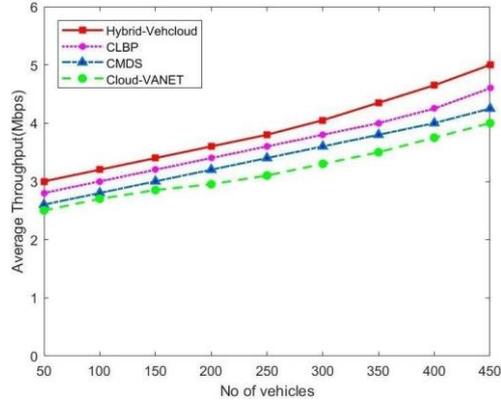

**Figure 1.7** Comparison of average throughput of Hybrid-Vehcloud with CLBP, CMDS, and Cloud-VANET approaches.

### 1.4.4 PERFORMANCE EVALUATION OF DFCV

DFCV is compared with three previous fog-based message dissemination schemes: 1) Named Data Networking (NDN) with mobility [31], 2) Fog-NDN with mobil- ity [32], and 3) PEer-to-Peer protocol for Allocated REsource (PrEPARE) protocols [41]. The results are discussed below:

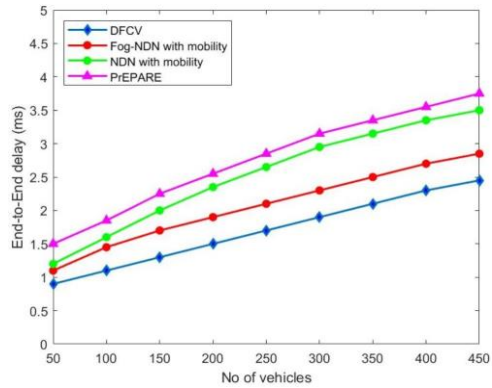

**Figure 1.8** Comparison of end-to-end delay of DFCV with NDN with mobility, Fog-NDN with mobility, and PrEPARE approaches.

**End-to-end delay:** In DFCV, as fog nodes located at the proximity of users, it reduces the time taken for an initial setup across a network from source to destination and disseminate the messages much quicker than existing approaches such as fog-NDN with mobility, NDN with mobility, and PrEPARE protocols. The end-to-end delay is calculated against the number of vehicles and is observed to be lower at all vehicle densities, represented in Fig. 1.8.

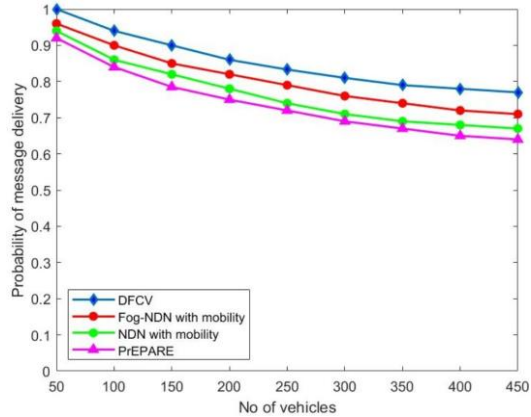

**Figure 1.9** Comparison of probability of message delivery of DFCV with NDN with mobility, Fog-NDN with mobility, and PrEPARE approaches.

**Probability of message delivery:** The probability of message of delivery of DFCV was observed to higher like Hybrid-Vehcloud (Section 1.4.3.) as DFCV also provides guaranteed message delivery at all vehicle densities. It is calculated against the number of vehicles, represented in Fig. 1.9.

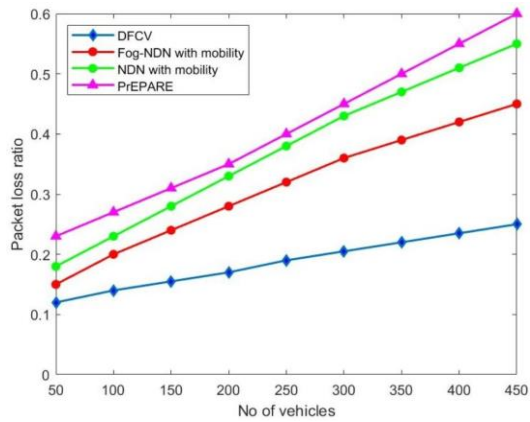

**Figure 1.10** Comparison of packet loss ratio of DFCV with NDN with mobility, Fog-NDN with mobility, and PrEPARE approaches.

**PLR:** To observe the ratio of number of lost packets in a network before reaching the destination, we performed this experiment at a time interval (t) and observed that the PLR of DFCV approach is lower at high vehicle densities. PLR increases slightly as the number of users increases in the system, as shown in Fig. 1.10. It is due to the additional packets generated being more likely to encounter another packet and resulting in a collision.

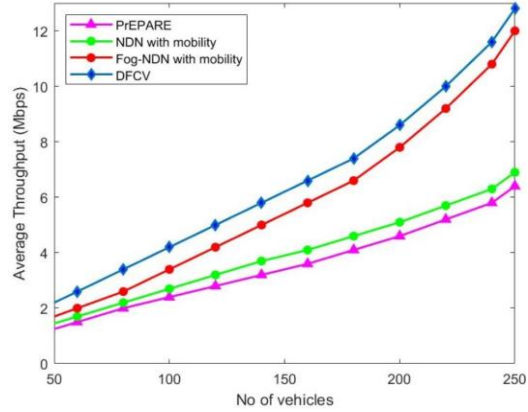

**Figure 1.11** Comparison of average throughput of DFCV with NDN with mobility, Fog-NDN with mobility, and PrEPARE approaches.

**Average Throughput:** The average throughput of DFCV is compared with fog-NDN with mobility, NDN with mobility, and PrEPARE protocols. It is calculated against the number of vehicles, represented in Fig. 1.11. The number of messages disseminated across a network increases as the number of vehicles increases in a system. As DFCV, provides guaranteed message delivery at high vehicle densities, the average throughput of DFCV is observed to be higher compared to fog-NDN with mobility, NDN with mobility, and PrEPARE protocols at all vehicle densities.

## 1.5   COMPARISON OF VEHICULAR FOG COMPUTING AND VEHICULAR CLOUD COMPUTING

This section compares and contrasts between the benefits from the vehicular fog computing against the vehicular cloud computing and vice versa.

### 1.5.1   ADVANTAGES OF VEHICULAR FOG COMPUTING OVER VEHICULAR CLOUD COMPUTING

1. Low latency: Fog nodes are formed at the proximity of end-users. The processing of data takes place on the edge, much closer to the vehicles, resulting in rapid transmission of messages to other vehicles and fixed infrastructures like RSUs, etc.

2. Better QoS: Vehicular fog computing provides better data transmission rates with minimum response time compared to vehicular cloud computing at high vehicle densities. Thus, vehicular fog computing provides better QoS than vehicular cloud computing.

3. Network efficiency and energy consumption: Unlike vehicular cloud computing, vehicular fog computing avoids the back and forth transmission between cloud servers. Thus, the bandwidth utilization and energy consumption of vehicular fog computing are much lesser compared to vehicular cloud computing.

4. Improved agility of services: The rapid development of vehicular fog computing allows the users to customize the applications nearer to them instead of sending the changes to vehicular cloud servers and waiting for the response from the cloud servers.

5. Deployment cost: The deployment cost of vehicular fog computing is very less compared to vehicular cloud computing. Any real-world objects that have the following properties such as, network connectivity , storage , and computing can become a fog node . Thus, the deployment cost of fog is lesser than vehicular cloud computing.

### 1.5.2 ADVANTAGES OF VEHICULAR CLOUD COMPUTING OVER VEHICULAR FOG COMPUTING

1. Storage : The vehicular cloud is centralized, offers wide storage space compared to vehicular fog computing. The users can deploy space constraint applications in the cloud server rather than fog nodes. Fog nodes provide limited storage space to the users, which cannot be used applications that require more storage space.

2. Resource management: Vehicular cloud computing manages and dynamically change the cloud resources based on departure and arrival of vehicles. Each vehicle negotiates the level of resource sharing directly. Whereas, efficient utilization of resources is being considered as a major research area in vehicular fog computing.

3. Service: Vehicular cloud computing is popular in providing services on pay per use to the user based on demand. Vehicular cloud computing offers three types of services such as Network as a Service (NaaS), Storage as a Service (SaaS), and Cooperation as a Service (CaaS). Whereas, vehicular fog computing does not provide services on pay per use to the users.

## 1.6  CONCLUSION

In this book chapter, we provide an extensive overview of message dissemination techniques in connected vehicles including, challenges and various scenarios involved in message dissemination based on two major classifications: vehicular cloud-based message dissemination, and vehicular fog-based message dissemination. We discussed the working principle of two reliable and efficient message dissemina-tion frameworks: Hybrid-Vehcloud and DFCV to help the readers better understand the fundamentals of connected vehicles, communication techniques , and various scenarios to be considered while disseminating messages. Above all, we have illustrated the performance of various message dissemination protocols including, Hybrid-Vehcloud and DFCV based on performance metrics like end-to-end delay , PLR , average throughput , and probability of message delivery .

## 1.7 FUTURE DIRECTIONS

In the future, the dissemination of messages in connected vehicles will extend using machine-learning techniques. Machine learning can be implemented among connected vehicles using various techniques including, decision trees, decision tree ensembles, artificial neural networks, and k-nearest neighbors. The use of machine learning in broadcasting messages will monitor the path periodically and transmit the message in an optimized route. This will increase the performance of the system in an efficient manner. Furthermore, integration of VANETs and 5-G will lead to significant enhancements and efficiency in vehicular communications.

With an increase in the number of vehicles, a huge volume of data being generated. These data should be monitored, analyzed, and managed properly to reduce storage and bandwidth consumption. Advanced data processing and data mining techniques can be applied to handle the large volume of data disseminated from the active participants in a connected vehicular environment. Security and privacy are another major concern for VANETs. The internet connecting to a large number of vehicles and infrastructures suffering from various network attacks like Denial of Service (DoS) attack, timing attack, Sybil attack, false information attack, black hole attack, tunneling attack, etc. Intrusion Detection System (IDS) can be implemented to detect network attacks caused by the compromised node on the network. The objective of IDS is to protect the network from network-based threats. IDS is located at some special nodes called monitoring nodes. The deployment of the monitoring nodes differs depending on the protocol type and the architecture of the IDS. Furthermore, new secure communication protocols must be investigated under various networking conditions.

# Index